\documentclass[twocolumn,epsfig,aps]{revtex4}
\begin{document}
\draft
\title{Encoding a logical qubit into physical qubits}
\author{B. Zeng,$^{1,\dag}$ D. L. Zhou,$^2$ Z. Xu,$^{1}$
C. P. Sun,$^2$ and L. You$^{2,3}$}
\address{$^1$Department of Physics, Tsinghua University,
Beijing 100084, P.R. China}
\address{$^2$Institute of Theoretical Physics, The Chinese
Academy of Sciences, Beijing 100080, P. R. China}
\address{$^3$School of Physics, Georgia Institute of Technology,
Atlanta, Georgia 30332, USA}
\date{\today}

\begin{abstract}
We propose two protocols to encode a logical qubit into physical
qubits relying on common types of qubit-qubit interactions in as
simple forms as possible. We comment on its experimental
implementation in several quantum computing architectures, e.g.
with trapped atomic ion qubits, atomic qubits inside a high Q
optical cavity, solid state Josephson junction qubits, and
Bose-Einstein condensed atoms.
\end{abstract}
\maketitle

The extreme power of quantum information processing originates
from the coherence properties of quantum states, manifested in
terms of entanglement between different local (physical) qubits
\cite{Ni}. In recent years, a variety of physical qubits have been
realized, and controlled qubit-qubit interactions demonstrated,
pointing to a promising future for large scale quantum computing
technology. However, the much studied environment induced
decoherence and dissipation will inevitably destroy quantum
coherence, leading to unavoidable errors of physical qubit based
quantum information.

The most remarkable difference between quantum and
classical errors is due to the fact that quantum errors
are continuous, thus can not be simply corrected with known
error correcting schemes from classical information theory.
In 1995, Shor made the important discovery that
9-qubit based logic qubit can correct
arbitrary quantum errors, despite their continuous nature
\cite{sho95}. Following this landmark achievement,
quantum error codes \cite{CS,Ste} and fault
tolerant quantum computing \cite{sho96,kit97,cpz96} have been
extensively studied, significantly boosting the confidence
in practical quantum computation.

The continued development of quantum information science
highlights the importance of mapping physical qubits into logic
qubits. In an error correcting procedure as we now understand,
encoding a logic qubit into physical qubits is the important first
step that introduces extra freedom to store less information in
order to recover from errors. In addition, the logic qubit basis
state can be chosen to be a decoherence free subspace, offering
the potential advantage of completely eliminating environment
induced decoherence \cite{zr98,lcw98,wineland}. Of all
architectures considered for experimental quantum computation, the
encoding of logic qubits into physical qubits has only been
accomplished, and in fact, used extensively in ensemble based NMR
systems. A simpler version, encoding a qubit into a two
dimensional decoherence free subspace was recently demonstrated
with two trapped ions \cite{wineland}.

In this study, we investigate the proper interactions needed for
realizing efficient encoding of logic qubits. We suggest two
protocols that accomplish the intended encoding of an unknown
qubit relying on simple interactions as commonly used in the
generation of maximally entangled states of various two state
systems, e.g., Bose-Einstein condensed atoms \cite{You}, trapped
atomic ions \cite{Mol}, cavity QED systems \cite{cqed}, and solid
state Josephson junctions \cite{jc2}. In what follows, we will
first detail our protocol for a model of many qubits. This is
followed by more specific discussions of several physical systems
where our model can be applied; we then conclude with a brief
summary of our study.

We start with a statement of the problem being considered here in
this article, i.e. encoding of an unknown logic qubit into
physical qubits. The unknown qubit takes the general form
$\alpha|0\rangle+\beta|1\rangle$ with coefficients $\alpha$ and
$\beta$ and orthogonal computation basis states $|0\rangle$ and
$|1\rangle$. The logic qubit typically involves several qubits and
is capable of correcting certain types of error. As a simple
example, we first consider an encoded logic qubit in terms of the
repeated state (of $N$ additional qubits) that reads
$\alpha|\underbrace{0,\cdots,0}_{N+1}\rangle
+\beta|\underbrace{1,\cdots,1}_{N+1}\rangle$. This encoding
clearly can correct bit flip errors as long as the number of
flipped qubits is less than half of the total (N+1), despite its
extreme sensitivity to phase errors. In the end, we will also give
explicit steps for constructing the 9-qubit Shor code \cite{sho95}
that corrects all known types of 1-qubit errors.

Our initial logic qubit thus uses a total of $N+1$ physical qubits
with the first physical qubit being the given (unknown) qubit. Without loss of
generality, the initial state of our system is thus
\begin{eqnarray}
|\psi(0)\rangle=(\alpha|0\rangle+\beta|1\rangle)
\otimes|\underbrace{0,\cdots,0}_N\rangle, \label{eq1}
\end{eqnarray}
with $N$ assumed even. Relying on pair wise interactions, a
straightforward encoding would involve the use of $N$ CNOT gates
with the first physical qubit as control and each of reminder $N$ qubit
as target. This strategy of using repeated CNOT gates is clearly not the most
economical protocol for our intended encoding. As our investigation will show,
there exist more efficient protocols that use common types of
qubit-qubit interactions in a collective way to accomplish the
encoding in as simple forms as possible.

One of the interaction terms, the collective Hamiltonian we intend
to use for encoding, is $H=uJ^2_x$ with
$J_x=\hbar\sum_{j=2}^{N+1}\sigma_x^{(j)}/2$ (or equivalently
$J_y^2$ or $J_z^2$ as in Ref. \cite{You}). $\sigma_x^{(j)}$ is the
Pauli matrix of the jth qubit. Such a Hamiltonian is symmetric and
takes a simple collective form. Its importance has been recognized
by several groups in the study of generating maximally entangled
or spin squeezed states \cite{You,Mol}. We assume the interaction
strength $u$ is controllable, i.e. can be turned on and off to
affect the dynamics of the $N$ added qubits. In the second
quantization language, with $a(a^{\dag})$ and $b(b^{\dag})$ as the
annihilation (creation) operators for the qubit state $|0\rangle$
and $|1\rangle$, we find the expression
$J_x=(b^{\dag}a+a^{\dag}b)/2$, completely symmetric with each of
the $N$-qubits.

Previous studies \cite{You,Mol} have shown at time
$\tau_1=\pi/2u$, the Hamiltonian $uJ^2_x$ will evolve the initial
$|\phi_0(0)\rangle=|\underbrace{0,\cdots,0}_N\rangle$ into a
maximally entangled state, or a GHZ state of the $N$ qubits
\begin{eqnarray}
|\phi_0(\tau_1)\rangle=\frac{1}{\sqrt{2}}
[e^{-i\frac{\pi}{4}}|\underbrace{0,\cdots,0}_N\rangle
+e^{i\left(\frac{\pi}{4}+\frac{N}{2}\pi\right)}|\underbrace{1,\cdots,1}_N\rangle].
\label{eqe0}
\end{eqnarray}
By symmetry, the initial state
$|\phi_1(0)\rangle=|\underbrace{1,\cdots,1}_N\rangle$ also evolves
into a maximally entangled state
\begin{eqnarray}
|\phi_1(\tau_1)\rangle=\frac{1}{\sqrt{2}}
[e^{i\left(\frac{\pi}{4}+\frac{N}{2}\pi\right)}|\underbrace{0,\cdots,0}_N\rangle
+e^{-i\frac{\pi}{4}}|\underbrace{1,\cdots,1}_N\rangle].
\label{eqe1}
\end{eqnarray}
At time $\tau_2=2\pi/u$, i.e. a period of evolution from the
initially separable states $|\phi_{0/1}(0)\rangle$ to a maximally
entangled states $|\phi_{0/1}(\tau)\rangle$ and then back to the
initial separable states, we end up with
$|\phi_0(\tau_2)\rangle=|\underbrace{0,\cdots,0}_N\rangle$ and
$|\phi_1(\tau_2)\rangle=|\underbrace{1,\cdots,1}_N\rangle$.

Based on the above dynamics of $uJ_x^2$ periodically creating
maximally entangled states from separable states, we devise the
following protocol for efficiently encoding an unknown logical
qubit into physical qubits. We start with the initial state as
given in Eq. (\ref{eq1}).

First, we turn on the $uJ^2_x$ interaction of the $N$ qubits [from
the 2nd to the $(N+1)$-th]. After a duration of $\tau_1$, the
complete quantum state becomes
\begin{eqnarray}
|\psi(\tau_1)\rangle=(\alpha|0\rangle+\beta|1\rangle)
\otimes|\phi_0(\tau_1)\rangle.
\end{eqnarray}

We now turn off the $uJ^2_x$ interaction and turn on another
interaction $\sigma_z^{(1)}\sigma_z^{(2)}$ between the first physical
qubit and another one selected arbitrarily from the other $N$
qubits. This second interaction is necessarily similar to
that required in the detection of the maximally entangled states
as it provides individual addressibility for at least one of the
additional $N$ qubits, as discussed by Itano {\it et al.}
\cite{itano}. After an additional duration $\tau$, this second
interaction leads to $|00\rangle\rightarrow|00\rangle$,
$|01\rangle\rightarrow|01\rangle$,
$|10\rangle\rightarrow|10\rangle$, and $|11\rangle\rightarrow
-|11\rangle$, i.e. a simple phase gate between the physical qubit
and an arbitrary one of the $N$ additional qubits. In fact, other
gates, such as a CNOT can also be used here as a substitute to
effect our protocol, although additional single bit rotations may
be needed in this case. The state of our system then becomes
\begin{eqnarray}
|\psi(\tau_1+\tau)\rangle=
\alpha|0\rangle|\phi_0(\tau_1)\rangle-i(-1)^{N/2}\beta|1\rangle|\phi_1(\tau_1)\rangle.
\end{eqnarray}

We then turn off the $\sigma_z^{(1)}\sigma_z^{(2)}$ interaction
and turn back on the $uJ^2_x$ interaction for the last $N$ qubits
again and after a time $\tau_2-\tau_1$ we finally obtain
\begin{eqnarray}
|\psi(\tau_2+\tau)\rangle=
\alpha|\underbrace{0,\cdots,0}_{N+1}\rangle
-i(-1)^{N/2}\beta|\underbrace{1,\cdots,1}_{N+1}\rangle,
\end{eqnarray}
i.e., exactly what the encoding task calls for apart from a
known relative phase factor $-i(-1)^{N/2}$\ between the logic
qubit basis states $|\underbrace{1,\cdots,1}_{N+1}\rangle$ and
$|\underbrace{0,\cdots,0}_{N+1}\rangle$. This phase factor can be
easily eliminated using single qubit rotations without knowing the
coefficients $\alpha$ and $\beta$.

Now we turn to the second protocol for the same encoding
that uses only the $uJ_{x}^{2}$ type interaction
and single qubit measurement. The required addressibility of a
particular qubit from the N appended qubits to effect the
$\sigma_z^{(1)}\sigma_z^{(2)}$ interaction as in the first protocol
is now eliminated. Starting from the same initial state
(1), we now assume $N$ odd for simplicity with $N+1$ being
an even number. This second protocol is possible because of
an important observation that the Hamiltonian
$H=uJ_x^2$ also evolves the initial state
$|\underbrace{1,0,\cdots,0}_{N+1}\rangle$ to a maximally entangled
state $\frac{1}{\sqrt{2}}e^{-i\pi/4}(|\underbrace{1,0,\cdots,0}_{N+1}\rangle
+i(-1)^{N/2}|\underbrace{0,1,\cdots,1}_{N+1}\rangle)$.
Therefore, the second protocol consists of the
following: First we apply the interaction $uJ_{x}^{2}$ to all
the $N+1$ qubits of $|\psi(0)\rangle$. This leads to
the time evolved state (apart from an overall phase) at time
$\tau_1=\pi/2u$
\begin{eqnarray}
|\psi{(\tau_1)}\rangle&=&\alpha(|\underbrace{0,0,\cdots,0}_{N+1}\rangle
+e^{i\phi}|\underbrace{1,1,\cdots,1}_{N+1}\rangle)\nonumber\\
&&+\beta(e^{i\phi}|\underbrace{1,0,\cdots,0}_{N+1}\rangle
+|\underbrace{0,1,\cdots,1}_{N+1}\rangle)\nonumber\\
&=&|0\rangle(\alpha|\underbrace{0,\cdots,0}_{N}\rangle
+e^{i\phi}\beta|\underbrace{1,\cdots,1}_{N}\rangle)\nonumber\\
&&+|1\rangle(\alpha e^{i\phi}|\underbrace{1,\cdots,1}_{N}\rangle
+\beta|\underbrace{0,\cdots,0}_{N}\rangle),
\end{eqnarray}
with $\phi=(N+1){\pi/2}$.

Then we measure the first physical qubit in the $\sigma_z$ basis: if we get
$|0\rangle$, the encoding task is accomplished preceded by a
single qubit rotation to get rid of the $e^{i\phi}$; if we get
$|1\rangle$, then another $uJ_x^2$ interaction for a period
$\pi/u$ can be applied to the other $N$ physical qubits to exchange the
place of $|0\rangle$ and $|1\rangle$. This exchange of basis
states can also be achieved using simultaneous single bit Rabi
oscillations. As before, single qubit rotations are needed to get
rid of the phase factor $e^{i\phi}$ and to take into account the
fact that for $N$ odd, a linear interaction $J_x$ is also needed
\cite{Mol}.

The above two theoretical protocols have three distinct features
of the encoding procedure. First, all operations are independent
of the details of the initial physical state, i.e., an unknown
logic state can be encoded into corresponding physical qubits.
Second, the complexity of the encoding operation is not enhanced
with the increase of the number of physical qubits in one logic qubit.
Third, the model interactions used are of relative simple forms,
and have already been engineered in several real physical systems.

We note that
the sensitivity to phase errors of the above logic qubit can
be easily overcome with a basis rotation.
By applying ($\pi/2$) Rabi pulses to all qubits, we effectively
affect the Hadamard gate to all qubits, leading to a logic qubit
$\alpha|\underbrace{+,\cdots,+}_{N+1}\rangle
+\beta|\underbrace{-,\cdots,-}_{N+1}\rangle$, that clearly
can correct phase errors as long as the number of
error qubits is less than half of the total (N+1). In the following,
we discuss the various implementations
of the above encoding protocols within several architects
of experimental quantum computation efforts.

Firstly, trapped atomic ions has emerged as one of the most attractive
prospective quantum computer architectures, setting the standard
for rudimentary quantum logic operations on several qubits
\cite{czion,blatt,nist}.
The strong Coulomb interaction between qubits is internal state
independent, and permits quantum information
to be controllably transferred and entangled between
several neighboring ions sharing the same quantum data bus of
a collective vibrational mode. When irradiated by external
laser fields, both types of interactions as required above
for the encoding protocol have been demonstrated before,
in particular, the $uJ_x^2$ type interaction was used to
entangle 4 separate ions without individual addressing
from the external laser fields \cite{Mol,nist4} and the
$\sigma_z^{(1)}\sigma_z^{(2)}$ is equivalent to a CNOT operation
between two neighboring ions, where a variety of equivalent
protocols have been also engineered \cite{blatt,nist}. Furthermore,
within this system, we can even imagine the physical qubit to be
of a different type of ion from the remainder N ion qubits,
as recently shown in the experimental
work on sympathetic cooling of ions \cite{nists}.
Thus its state can be prepared and individually addressed,
resolved, or detected using its frequency space (internal state)
selectivity,
eliminating the difficult task of focusing onto the tight
space of each ion qubit in an array of trapped ions.

As an application of the encoding protocol suggested
above, we show that the Shor nine qubit code \cite{sho95}
can be directly implemented. Starting with the (physical)
qubit $\alpha |0\rangle+\beta|1\rangle$, we append
8 additional ion qubits from the same trapped ion array.
On application of our encoding protocol, the logic
qubit simply becomes
\begin{eqnarray}
\alpha |0,0,0\rangle|0,0,0\rangle|0,0,0\rangle
+\beta|1,1,1\rangle|1,1,1\rangle|1,1,1\rangle,
\label{oen}
\end{eqnarray}
where the ions are grouped into triplets in
preparation for the Shor coding scheme. Now for each
group of the three ion qubits, we turn on the
$uJ_x^2$ interaction
as induced from two Raman matched external laser
fields \cite{Mol}. After a duration of $\tau_1$,
and from Eqs. (\ref{eqe0}) and (\ref{eqe1}), we
accomplish the Shor coding
\begin{eqnarray}
&&\alpha[
{1\over \sqrt 2}(|0,0,0\rangle-|1,1,1\rangle)]^{\otimes 3}\nonumber\\
+&&\beta [
{1\over \sqrt 2}(|0,0,0\rangle+|1,1,1\rangle)]^{\otimes 3},
\end{eqnarray}
upto an additional phase factor between states
$|0,0,0\rangle$ and $|1,1,1\rangle$.
(Note however that single qubit operations are also needed
in addition to $uJ_x^2$
as $N=3$ being an odd number \cite{Mol}).

Secondly, the recent success of quantum computation with
superconducting Josephson junctions has raised the
interests of the solid state device and material science
community. While it remains completely unclear
what types of physical systems the future's quantum
logic devices will be embedded in, the hardware industrial
base of the Information Technology does lend support
for a solid state based architecture \cite{jc}.
Since the demonstration of quantum coherent two states
in terms of a charge qubit, flux qubit, and the
charge-phase qubit, significant gains have been
accomplished in this direction \cite{jc1}. Already two independent
groups have reported the observation of entanglement
and correlation between two separate Josephson junction
qubits \cite{cai,lobb}. A number of theoretical proposals have
suggested system implementations for the required
interactions $u J_x^2$ and $\sigma_z^{(1)}\sigma_z^{(2)}$
or its equivalent forms \cite{jc2,jc}.

Thirdly, cavity QED based quantum computing architecture occupies
an unique place in quantum information science because it allows
for coherent exchange
of quantum information between material (atomic) and
photonic qubits \cite{cqed,kimble,mqed}. When carried by photons,
qubits enjoy such a weak coupling to the environment
that they can easily be distributed over very large distances
through optical fibers or free space, offering prospects
for distributed quantum computation and quantum communications.
In currently pursued cavity QED systems, trapped ion arrays
or neutral atoms intersect a high Q optical cavity, leading
to a scalable geometry with several atomic qubits that
can collectively share the same quantized cavity mode,
a quantum data bus of the electromagnetic field \cite{cqed,cqed1}.
A mapping to an equivalent ion trap system can be easily established,
leading to the recognition that both types of interactions
needed for our encoding protocol
can also be realized provided several experimental challenges,
especially the precise localization of each atom's position
inside the small cavity, can be met \cite{you1,zheng}.

Finally, the tremendous success of trapped atomic quantum gases
has also impacted atomic physics based quantum computing
efforts. Atomic Bose-Einstein condensates display remarkable
phase coherence and in addition, have demonstrated controllability
over atomic interactions and spatial arrangements \cite{zolc,expc}.
While it is not entirely clear how to perform (distinguishable)
qubit based quantum computation with Bose condensed (identical)
atoms, it is evident that the collective interaction $uJ_x^2$
arises naturally in all multi-component condensates as well as
spatially disconnected condensates \cite{You}. The second type
of interaction is more difficult to imagine at first sight,
because being a system of identical particles (bosons), it becomes
impossible to identify a 2nd qubit for an interaction
$\sigma_z^{(1)}\sigma_z^{(2)}$ with the first physical qubit.
Nevertheless, one might imagine loading an atomic condensate
into an optical lattice, and by mechanically moving the
trapped physical qubit into a particular lattice site,
one could engineer $\sigma_z^{(1)}\sigma_z^{(2)}$ using
the exchange interaction \cite{duan}, or
perhaps with atomic dipolar interaction as was suggested
in Ref. \cite{you2}. The lattice based atom interaction
among different sites, can also
be easily arranged to be of the form $uJ_x^2$ without
knowing which site is empty or filled \cite{mol1}.

Before concluding, we hope to address two important questions
regarding our protocol for encoding a logic qubit. First, can our
protocol be made fault tolerant? The simple answer is probably a
yes, although sacrificing simplicity and efficiency of the
collective interaction. Our protocol aims at an economical
approach to encode an unknown logic qubit into physical qubits,
not at the more complex question of fault tolerant quantum
computation in the current study \cite{shor96}. Yet the simple
version of fault tolerance with concatenated codes \cite{preskill}
can be similarly implemented as we demonstrated above that our
logic qubit basis states can be easily transformed to the Shor
code, which is capable of correcting arbitrary single bit errors.
Thus, our code can be made fault tolerant at least through
concatenation if the errors are not collective. If, however, the
errors are correlated in the symmetric $J_x^2$ interaction, then
potential problems do arise that will become subjects of our
future study. Second, does the requirement of error correction
negate the benefit in complexity that is at the heart of our
protocol? Our encoding protocol leads to the simple logic qubit
Eq. (\ref{oen}) capable of correcting only ``discrete" bit-flip
errors, not the more general ``continuous" errors. We do not
believe this is a draw as experimental effort in logic codes is
still at its infancy, and any proof of principle protocol will
benefit its development. Furthermore, as we have also shown, that
with a few extra steps and at the cost of requiring single bit
addressing, we arrive at the more advanced Shor code, capable of
correcting all ``continuous" single bit errors. Including the
overhead of these extra steps, our protocol still represents a
significant reduction in complexity for constructing the Shor code
as compared to the more straightforward approach with individual
quantum gates.

In conclusion, we have suggested two protocols for encoding a
logical qubit into physical qubits. The first one uses two types
of interactions: a collective symmetric interaction $uJ_x^2$ among
the appended qubits and an individually resolved qubit interaction
$\sigma_z^{(1)}\sigma_z^{(2)}$ between the physical qubit and any
one of the appended qubits. The second one requires only a
collective symmetric interaction $uJ_x^2$ and single qubit
measurement. We have aimed for the most economical types of
interactions, being simple and as symmetric as possible. While the
particular encoding scheme does not seem to offer much advantage
over the more sophisticated error correcting schemes, e.g. 5-qubit
and 7-qubit codes as discussed in theory, we have shown the our
code directly leads to the realization of the 9-qubit Shor code.
In addition, our work suggests a new milestone to calibrate
various experimental efforts in quantum computation; we note that
previous milestones have been 1), the demonstration of a physical
qubit, and 2) the demonstration of qubit-qubit entanglement; we
hope our work will shed light on the continued progress of quantum
information science \cite{fen}.

We thank Dr. E. Knill for helpful communications.
This work is supported by CNSF.
L. You also acknowledges support from NSF.

\end{document}